\pdfoutput=1
\documentclass[pre,aps,twocolumn,showpacs,floatfix,nofootinbib,superscriptaddress]{revtex4-2}
\usepackage{subfigure}
\usepackage{amssymb}
\usepackage{amsfonts}
\usepackage{amsmath}
\usepackage{amsthm}
\usepackage{epsfig}
\usepackage{graphicx}
\usepackage{cancel}
\usepackage{xcolor}
\usepackage{color}
\usepackage[latin1]{inputenc}
\usepackage[hidelinks,unicode=true]{hyperref}
\usepackage{comment}
\usepackage[normalem]{ulem}
\hypersetup{colorlinks=true,
 	linkcolor=blue,
 	urlcolor=blue,
 	citecolor=blue,
 	pdfhighlight=/N
 }

\renewcommand{\vec}{\mathbf}

\begin{document}
\title{ Scaling, fractal dynamics, and critical exponents in the equilibrium phase transition} 
\author{Adauto F. Souza}
\email{adauto.souza@ufrpe.br  }
\address{Departamento de F\'{\i}sica, Universidade Federal Rural de Pernambuco, Recife - PE, 52171-900, Brazil}
\author{Henrique A Lima}
\email{Henrique_adl@hotmail.com}
\address{International Center of Physics, Institute of Physics, University of Brasilia, 70910-900, Brasilia, Federal District, Brazil}
\author{Anderson L. R. Barbosa}
\email{anderson.barbosa@ufrpe.br}
\address{Departamento de F\'{\i}sica, Universidade Federal Rural de Pernambuco, Recife - PE, 52171-900, Brazil}
\author{Fernando A. Oliveira}
\email{faooliveira@gmail.com}
\address{Departamento de F\'{\i}sica, Universidade Federal Rural de Pernambuco, Recife - PE, 52171-900, Brazil}
\address{International Center of Physics, Institute of Physics, University of Brasilia, 70910-900, Brasilia, Federal District, Brazil}
\begin{abstract}

Statistical methods are essential for understanding thermodynamic systems with many degrees of freedom.
For systems in equilibrium, a very useful method is that of correlation functions, which establish a correlation between a field $\phi(\vec{x})$ that depends on the spatial position $\vec{x}$, with itself (autocorrelation function) at another position $\phi(\vec{x}_0)$. Fisher [Journal of Mathematical Physics 5, 944322 (1964)] introduced the autocorrelation function for fluctuations of the order parameter, which has been an important mathematical tool for understanding the second-order phase transition in equilibrium. However, his analysis is restricted to a Euclidean space of dimension $d$, and an exponent $\eta$ is introduced to correct the spatial behavior of the correlation function in $T=T_c$. In a recent work, Lima {\it et al.} [Phys. Rev. E 110, L062107 (2024)] demonstrated that a modern fractional differential analysis is necessary for a complete description of the correlation function at $T_c$. 
In this study, we highlight the deep connection among scaling behavior, critical exponents, and fractal geometry. Our results provide a unified geometric interpretation of critical exponents and fractal dimensions, broadly applicable to thermodynamic phase transitions. However, the approach does not apply to topological phase transitions, which lack local order parameters and the associated scale-invariant fractal geometry. We verify its predictions for several cornerstone thermodynamic models --- the Ising, Potts, XY, and Heisenberg systems.

\end{abstract}
\maketitle
\section{Introduction}
\label{Int}
Phase transitions are typically described by symmetry breaking, which is reflected in the order parameter and the correlation length. This becomes particularly clear for a second-order phase transition, where the order parameter vanishes at the critical temperature $T_c$, recovering the symmetry, and the correlation length diverges, thereby exhibiting scale invariance.
Thus, it is a very broad phenomenon that appears in many branches of physics.
To explore critical phenomena, we use lattice models that enrich statistical physics and quantum field theory. Lattice models are usually defined in Euclidean integer dimension $d$, while critical exponents are usually given by fractal numbers. An important quantity to evaluate is the time-independent correlation function
\begin{equation}
\label{Cdef}
    C(r)=\frac{1}{L^d} \sum_j \phi(\vec{j}+\vec{r}) \phi(\vec{j}),
\end{equation}
where the sum is over a hypercube lattice of dimension $d$ and side $L$.  Here $\phi(i)$ is the fluctuation of the order parameter at the lattice site  $i$. This function can be evaluated numerically and hopefully analytically. For a second-order phase transition, a mean-field solution is given by
\begin{equation}
\label{Cd1}
    C(r)=C_1r^{2-d} \exp(-r/\rho),
\end{equation}
where $C_1$ is a constant, and $\rho$ the correlation length~\cite{Cardy96}.  For $T$ close to the critical temperature $T_c$, the  correlation length diverges as
\begin{equation}
\label{rhodivergence}
\rho \propto |T-T_c|^{-\nu}.
\end{equation}
In this limit, Eq. (\ref{Cd1}) clearly fails for $d=2$, showing a correlation function independent of $r$, which is physically inconsistent.
This mean field solution breaks down at the transition, necessitating an empirical correction through the introduction of Fisher exponent $\eta$ \cite{Fisher64}
\begin{equation}
\label{G2}
C(r) \propto
\begin{cases}
r^{2-d} \exp(-r/\rho) , &\text{ if~~ } r>\rho,\\
r^{2-d-\eta}, &\text{ if~~ }  r \ll\rho.\\
\end{cases}
\end{equation}
The homogeneous continuous medium function Eq. (\ref{G2}) is a formal solution of the equation
\begin{equation}
	\label{G}
	(-\nabla^2 +\rho^{-2})C(r)=\delta^{(d)}(r)  \ . 
\end{equation} 
Correlation functions represent a simple form of the fluctuation-dissipation relation (see reviews~\cite{Nowak22,GomesFilho25}) and are easily associated with susceptibility~\cite{Goldenfeld18,Lima24}.  
 The violation of the Fluctuation-Dissipation Theorem (FDT) is well known in the literature for structural glasses~\cite{Grigera99,Ricci-Tersenghi00,Crisanti03,Barrat98,Bellon02,Bellon06}, random exchange Heisenberg chain~\cite{Vainstein05}, proteins~\cite{Hayashi07}, KPZ dynamics~\cite{Kardar86,Rodriguez19}, mesoscopic radioactive heat transfer ~\cite{Perez-Madrid09,Averin10}, and ballistic diffusion as well \cite{Costa03,Costa06,Lapas07,Lapas08}. The FDT breaks down when ergodicity is violated \cite{GomesFilho21,Costa03,Wen23}. 
 
 The FDT fails for critical points in the sense that it cannot explain Eq.\ (\ref{G2}) for $ r \ll\rho$, where the $\eta$ exponent appears as a need for consistency with the exponents equalities, e.g., the Fisher scaling relation~\cite{Fisher64}
 \begin{equation}
\label{gamma}
\gamma =(2-\eta)\nu.
\end{equation}
  In fact, the response function in the Euclidean formulation yields $\eta=0$ in disagreement with reality, thus showing a failure of the FDT. 
  Previous results for growth dynamics showed that when moving to a fractal formulation, the FDT was recovered~\cite{GomesFilho21,Anjos21,Luis22,GomesFilho24}. 
Moreover, fractal dynamics at phase transitions shares important features with growth phenomena, the scale invariance~\cite{Family85,Rodrigues24} and the attempt to determine whether an upper critical dimension exists~\cite{Barabasi95,Rodrigues15,Alves16,Gomes19,Marcos26}.
As well, the critical exponent of the correlation length $\nu$ is related to the critical exponent of specific heat $\alpha$ through the hyperscale relation~\cite{Cardy96}
\begin{equation}
\label{alpha}
\alpha=2-d\nu,
\end{equation}
thus associating a thermodynamic variable with the divergence of the correlation length. In other words, a relationship between thermodynamics and geometry. 
  
Inspired by this, Lima {\it et al.}~\cite{Lima24}  showed that at $T=T_c$, where $ \rho \rightarrow \infty$, we have the self-organization of the system into fractal clusters and the correlation function $C(r)$ does not  satisfy Eq. (\ref{G}), since it is in a Euclidian space, rather it  obeys the equation
\begin{equation}
\label{G3}
(-\nabla^2)^\zeta C(r)=\delta^{d_R}(r),
\end{equation}
to reflect the dynamics being restricted to a fractal structure characterized by the Riesz fractal dimension\footnote[1]{Note that, to avoid confusion with the notation already present in the literature, we have changed the original notation from \cite{Lima24} to the notation used in \cite{Lima25,Carrasco26}. Our notation here follows the same definition as the last two.} $d_R$,  $d-1 \leq d_R \leq d$ in the Riesz fractional derivative~\cite{Muslih10} of order $\zeta$, $1/2 \leq \zeta \leq 1$. The solution of this fractional differential equation, associated with the proposed solution (\ref{G2}), leads us to \cite{Lima24}
\begin{equation}
\label{eta}
\eta=d-d_R=1-\zeta.
\end{equation}
Now the Fisher exponent is not an ad hoc imposition; it arises naturally from the solution of (\ref{G3}). In the renormalization-group framework, the anomalous dimension \(\eta\) --- which is non-zero for interacting critical theories in dimensions less than \(4\)  --- is obtained as the fixed-point value of the field anomalous dimension \cite{dupuis2026rg}.
Equation  (\ref{rhodivergence}) highlights the
scaling invariance at the transition, where the dynamics occur not in the Euclidean dimension of the lattice but in the fractal structure of clusters of all sizes. Furthermore, it shows the exponent $\eta$ as the deviation from the integer dimension to the fractal dimension, or, equivalently, from an integer derivative to a fractional Riesz derivative.  

The above result is not the only one to interpret the scaling behavior seen at critical points
as manifestations of fractal geometries. The Fortuin-Kasteleyn clusters~\cite{FORTUIN1972536} or Coniglio-Klein  droplets~\cite{Coniglio_1980} (FKCK) allowed to
unify geometry and thermodynamics. The fractal geometry of FKCK droplets is the fractal, percolating clusters that form exactly at the critical point. When such a mapping is available, it provides the deepest link between critical phenomena and fractal geometry, and it has paved the way for the most efficient computational algorithms for studying spin models~\cite{landau2021guide}.
The FKCK mapping has been investigated in particular in the context of percolation~\cite{Suzuki8, Kroger00, Devakul19}.
For example, the relation between the exponent of the order parameter $\beta$, and the fractal dimension of the ordered phase $d_f$ was  proposed by Suzuki, who also made a connection to the renormalization-group framework~\cite{Suzuki8}
\begin{equation}
\label{df}
d_f=d-\frac{\beta}{\nu}\;.
\end{equation}
If the critical point is the percolation one, the fractal structure is the infinite percolating cluster at the transition \cite{Grimmett06,Cruz23}.  In general, it is associated with the largest ordered cluster at the critical point \cite{Kroger00}. Combining Eqs. (\ref{gamma}), (\ref{alpha}), (\ref{eta}) and (\ref{df}), and using the Rushbrooke equality $\alpha+2\beta+\gamma=2$,  we obtain an important connection between $d_R$ and $d_f$ as~\cite{Lima24}
\begin{equation}
\label{dR2}
    d_R=2(d_f-1).
\end{equation}
The fractal dimension predicted by Eq. (\ref{df}), $d_f=15/8$ for the 2D Ising model,
coincides with that of Coniglio for the two-state Potts model based on a mapping to the two-dimensional Coulomb gas \cite{Coniglio89}. This yields $d_R=7/4$ and $\eta=1/4$, from Eqs.~(\ref{dR2}) and (\ref{eta}), respectively, in agreement with the known values of the Ising model in two dimensions. 
Subsequently~\cite{Lima25,Carrasco26}, these relationships were verified for non-integer dimensions in the range $1 \leq d \leq 4$. Furthermore, it was verified for disordered Ising systems~\cite{lima26}.

Note that according to Eq.~(\ref{eta}) the geometric limit $1/2 \leq \zeta \leq 1$, imposes a maximum value $\eta=1/2$. However, it is possible to overcome this limitation using a scaling method~\cite{Carrasco26}.

In this work, our goal is to verify the relationships (\ref{eta}) and (\ref{dR2}) in more complex systems, such as the Potts, XY, and Heisenberg models, and to explore how far our interpretation can take us. This work is organized as follows: Section \ref{II} presents our discussion of the two-dimensional Potts model, for which we can obtain exact results. In Section \ref{III}, we present results and discussion for the two-dimensional XY model, for which our relations do not apply, and the three-dimensional Ising, XY, and Heisenberg models, with results that are not exact, but highly precise. We conclude in Section \ref{IV}.

\section{The Potts model}\label{II}

The Potts model is a generalization of the Ising model to more than two components~\cite{Potts52,f.y.wu}. Although intensively studied, it is a very active field of modern research~\cite{Xu25}.
A simple  Hamiltonian for it can be expressed as
\begin{equation}
    H=- J \sum_{(i,j)}\delta_{s_i,s_j}-h_0  \sum_i \delta_{s_i,0}   
\end{equation}
where the double sum $(i,j)$ runs over nearest neighbor spins, $\delta$ is the delta of Kronecker, which is $1$ for $s_i=s_j$ and zero otherwise. $J$ is the spin-spin coupling constant and $h_0$ is an external magnetic field. Spins $s_i$ take on values $s_i=(0,1,2,\ldots, q-1)$. 
It is thus defined as the $q$ state Potts model.  It is now known that the Potts model is related to a number of outstanding problems in lattice statistics; the critical behavior has also
been shown to be richer and more general than that of the Ising model. For example,  $q=1$ is used for the study of percolation and complex networks, while for $q=2$ we have the Ising model, and for $q=4$ we have the Baxter-Wu model.
In subsequent efforts to explore its
properties, the Potts model has become an important
test ground for different methods and approaches
in the study of critical point theory~\cite{f.y.wu}.

An useful representation for the \(q\)-state Potts model local order parameter \(\vec{m}_i\) at site \(i\) is a \((q-1)\)-component vector. In this simplicial representation, each one of the $q$ states,
\(\sigma_i\), is mapped onto a unit vector
\(\vec{e}_{\sigma_i}\) pointing to one of the $q-1$ vertices of a simplex. Thus,
the order parameter at site \(i\) is given by \(\vec{m}_i = \vec{e}_{\sigma_i}\).
The global order parameter reads
\begin{equation}
    \vec{M} = \frac{1}{N} \sum_i \vec{m}_i,
\end{equation}
and the corresponding fluctuation field, such as in Eq.  (\ref{Cdef}), is 
\(\mathbf{\phi}_i = \vec{m}_i - \langle \vec{M} \rangle\).


\begin{table}
       	\begin{tabular}{c|ccccc}
		\hline	\hline
\hspace{4mm}	$q$\hspace{4mm} & \hspace{4mm} $0$ & \hspace{4mm} $1$ & \hspace{4mm} $2$ \hspace{4mm}   & \hspace{4mm} $3$ \hspace{4mm} & \hspace{4mm} $4$ \hspace{4mm}   \\ \hline
%
	$\beta$ & $1/6 $ & $ 5/36 $   & $1/8 $   &  $ 1/9 $ &  $ 1/12$  \\ 
		$\nu$ &  $ \infty$ &  $ 4/3 $   & $1$   &     $ 5/6 $    & $2/3$  \\ 
  		$d_f$ & $2$   &  $91/48$   & $15/8$   &     $ 28/15 $    & $15/8$ \\ 
		$d_R$ & $2$  & $43/24 $     & $7/4$   &      $ 26/15$   &  $7/4$  \\	
        $\eta$  & $0$  &   $5/24 $   & $1/4$   &     $ 4/15$     & $1/4$ \\ 
        $\zeta$ & $1$ &  $19/24 $   & $3/4$   &      $ 11/15$   &   $3/4$ \\
		 \hline
		 	\hline
	\end{tabular}	
	\caption{ { Values of critical exponents and fractal dimensions for  the 2D Potts model universality class for $q=0,1,2,3$, and $4$. $d_f$ from \cite{Coniglio89, f.y.wu}, which agree with Eq. (\ref{df}), and $d_R$ from  Eq. (\ref{dR2}), while $\eta$ from Eq.\ (\ref{eta}). Note the symmetric deviation for both the fractal dimension $d_R=d-\eta$ and the fractional derivative $\zeta=1-\eta$. }}
   \label{Table1}
\end{table}

\begin{table}
\begin{tabular}{ c | ccc }
\hline \hline
\hspace{4mm} Model      \hspace{4mm} &\hspace{4mm} Ising (3D)        \hspace{4mm} &\hspace{2.5mm} XY (2D)      \hspace{2.5mm} &\hspace{1mm} XY (3D) \hspace{4mm}\\ \hline
$\beta$ &   $0.326418(2) $   &  0      &  $0.3485(2),$ \\
$\nu$   &   $0.629971(4)$    &  $\infty$     &  $0.67155(27)$ \\
$d_{f}$ &    $2.481856(2)$   &   2   & $2.481(2)$ \\
$d_R$ &    $2.963713(5)$    &  2    &  $2.962(5)$\\
  $\zeta$ & $0.963713(5)$ &  $1 $   &  $0.9619(4)$  \\
  $\eta*$  &  $0.0362978(20)$   &  $1/4$   & $0.0380(4)$ \\ 
$\eta$  &   $0.036288(5)$ &  0  &  $0.0380(5)$ \\ \hline \hline
\end{tabular}
\caption{  Values of critical exponents and fractal dimensions for some models. $\beta$, $\nu$ and $\eta*$, (to distinguishe from our $\eta$) from literature. For  Ising 3D \cite{Chester20}, XY 2D \cite{Kosterlitz74} and XY 3D \cite{Campostrini02}. We obtain  $d_f$ from  Eq. (\ref{df}), and $d_R$ from  Eq. (\ref{dR2}), while $\eta$ from Eq.\ (\ref{eta}). }
   \label{Table2}
\end{table}

In Table \ref{Table1}, we present the values of $d_f$ and $d_R$ for the $2D$ Potts model universality class for $q=0,1,2,3$, and $4$, alongside the corresponding critical exponents. 
From \cite{f.y.wu} we obtain $\beta$ and $\nu$, and $d_f$ from Eq. (\ref{df}). The values agree with those of Coniglio~\cite{Coniglio89}. We note that the equation is fulfilled for all $q$ shown. From this, $d_R$, $\zeta$ and $\eta$ can be independently determined. Furthermore, note that the values of $\eta$ obtained this way agree with the values obtained exactly previously~\cite{f.y.wu}.
Notice that for $q=4$, $\beta$ and $\nu$ are different from the values for $q=2$. However, the ratio $\beta/\nu$ is the same. Thus, from Eq. (\ref{df}) we obtain the same $d_f$,  from Eq. (\ref{df}) and (\ref{dR2}) the same $d_R$  and the same $\eta$. The identical values of $\eta$ for $q=2$ and $q=4$ cannot be interpreted as a coincidence, but as an imposition of fractal geometry, which has the same $d_f$ and $d_R$ for both cases. Finally, note that our values of $\eta$ agree with the exact values reported in the literature. This reinforces the value of Eq. (\ref{eta}) for both the determination and the interpretation of the exponent $\eta$. 

\begin{table*}
\begin{tabular}{ c | cccc }
\hline \hline
\hspace{4mm} Ref.      \hspace{4mm} &\hspace{4mm} \cite{Guillou80} \hspace{4mm} &\hspace{4mm} \cite{Brezin85}       \hspace{4mm} &\hspace{4mm}  \cite{Holm93}      \hspace{4mm} &\hspace{4mm} \cite{Campostrini02}   \hspace{4mm}\\ \hline
$\beta$ &   $0.3645(25) $   &  $0.368(4) $   & 0.362(4)      &  $0.3689(3)$ \\
$\nu$   &   $0.705(3)$    &  $0.710(7) $   & $0.706(9)$     &  $0.71120(5)$ \\
$d_{f}$ &    $2.483(5)$   &  $2.482(11) $   &  $2.49(1)$  & $2.4813(5)$ \\
$d_R$ &    $2.97(1)$    &  $2.963(22) $   &  $2.97(2)$  &  $2.9626(9)$\\
  $\zeta$ & $0.965(10)$ &  $0.963(22) $   & $0.975(2) $   &  $0.9626(9)$  \\
  $\eta*$  &  $0.033(4)$   &   $0.040(3) $   & $0.027(2)$   & $0.0375(3)$ \\ 
$\eta$  &   $0.035(10)$ &   $0.037(40) $   & $0.025(2) $  &  $0.0374(9)$ \\ \hline \hline
\end{tabular}
\caption{  Values of critical exponents and fractal dimensions for Heisenberg 3D model. $\beta$, $\nu$ and $\eta*$, (to distinguishe from our $\eta$) from literature. We obtain  $d_f$ from  Eq. (\ref{df}), and $d_R$ from  Eq. (\ref{dR2}),while $\eta$ and $\zeta$ are from Eq.\ (\ref{eta}). }
   \label{Table3}
\end{table*}

\section{Beyond Potts Model}\label{III}

Following the order of complexity, we first consider the Potts model as a natural extension of the Ising model, then investigate the XY and Heisenberg models. The Ising 3D model is mentioned because its critical exponents are known with high precision.

\subsection{Ising 3D and XY Models}

Unlike the Potts model, the classical XY and Heisenberg models have continuous symmetry. Now the dynamical degrees of freedom consist of unity 2- and 3-component vectors \(\vec{S}_i\). The Hamiltonian for these model systems may be written
\begin{equation}
    H=- J \sum_{(i,j)} \left( \Delta S_i^z S_j^z + S_i^x S_j^x + S_i^y S_j^y\right),
\end{equation}
where \(J\) sets the overall energy scale, and \(\Delta\) is an anisotropy parameter. The XY model \cite{Kosterlitz74} is recovered setting \(\Delta = 0\), and the isotropic Heisenberg model \cite{PhysRevB.43.6087} corresponds to \(\Delta = 1\).
For $\Delta \ne 1$, the anisotropic Heisenberg system generally lies in a different universality class from the isotropic one \cite{PhysRevE.91.032146} (see subsection B); consequently, we will not discuss the anisotropic case here.

In Table \ref{Table2}, we present the values of $d_f$ and $d_R$ for the Ising 3D model and the XY model in 2 and 3 dimensions, along with the corresponding critical exponents. Here we use the notation $\eta*$ obtained from the literature to distinguish it from our $\eta$, Eq.\ (\ref{eta}). The values of $\beta$, $\nu$ and $\eta*$, are obtained from:   Ising 3D ref \cite{Chester20}, XY 2D ref \cite{Nijs83} and XY 3D \cite{Chester20}. We obtain  $d_f$ from  Eq. (\ref{df}), and $d_R$ from  Eq. (\ref{dR2}), while $\eta$ from Eq.\ (\ref{eta}). From this, $d_R$, $\zeta$ and $\eta$ can be independently determined. Furthermore, note that the values of $\eta$ obtained this way agree with the values of $\eta*$, within the error for the three-dimensional Ising and XY model.

An intriguing result emerges for the two-dimensional XY model, which does not exhibit Mermin-Wagner symmetry breaking, but possesses a BKT (Berezinskii-Kosterlitz-Thouless) topological transition~\cite{Devakul19,Kosterlitz73,Kosterlitz74,Kosterlitz17nobel,Gong19}. The $\eta$ parameter behaves as
\begin{equation}
\label{etaR}
\eta(T) =
\begin{cases}
0 , &\text{ if~~ } T < T_{BKT},\\
1/4 , &\text{ if~~ } T = T_{BKT},\\
\infty , &\text{ if~~ } T > T_{BKT},\\
\end{cases}
\end{equation}
Where $ T_{BKT}$ is the topological transition temperature. The main characteristic is the universal value of $\eta=1/4$  which, since it is not an order-to-disorder transition, does not agree with our result. Curiously, under this extreme condition, our approach predicts flat surfaces with $d_r=d_f=d=2$ and a return to the Euclidean mean-field results, $\eta=0$ and fractional derivative order $\zeta=1$.
Consequently, our formulation does not apply to topological phase transitions.

\subsection{Heisenberg Model}
The two-dimensional Heisenberg model represents an even more restrictive case than the two-dimensional XY model,  that is, it does not even present a topological transition. Consequently, we will not consider it here.

In Table \ref{Table3}, we present the values of the fractal dimensions $d_f$ and $d_R$ for the three-dimensional Heisenberg model, together with their corresponding critical exponents. Here we use the notation $\eta*$ obtained from the literature to distinguish it from our $\eta$, Eq.\ (\ref{eta}). The values of $\beta$, $\nu$, and $\eta*$ were obtained from the cited references. We obtain  $d_f$ from  Eq. (\ref{df}), and $d_R$ from  Eq. (\ref{dR2}), while $\eta$ from Eq.\ (\ref{eta}). From this, $d_R$, $\zeta$ and $\eta$ can be independently determined. Furthermore, note that the values of $\eta$ obtained in this way agree with the values of $\eta*$, within the error margins.

\section{Concluding remarks and perspectives}\label{IV}

We have introduced a new fractal structure associated with the dynamic correlations present at equilibrium critical points. This structure is characterized by a correlation fractal dimension. We relate this dimension to the Fisher exponent $\eta$, which characterizes the decay of the correlation function near criticality, via Eq.\ (\ref{eta}). Our approach replaces the traditional correlation function Eq.\ (\ref{G}) with a version based on the Riesz fractional derivative Eq.\ (\ref{G3}) at the critical point. The Riesz fractional derivative is a mathematical tool that generalizes differentiation to non-integer orders, capturing the fractal nature of correlations. This equation defines the fractal subspace determined by the correlations. Using this substitution restores the form of the correlation function at the critical point. Therefore, the Fisher exponent in the correlation function $C(r)$ indicates how much the correlation fractal dimension deviates from the system's integer spatial dimension. Both $\zeta$, a parameter controlling the decay rate, and $d_R$, the Riesz fractal dimension or correlation fractal dimension, fall within previously established bounds: $0<\zeta<1$ and $d-1 \leq d_R \leq d$.


Using precise estimates of previously published critical exponents, we obtained the fractal dimensions $d_R$ and $d_f$, as well as the Fisher exponent $\eta$ for the two-dimensional Potts model (Table 1, exact results). Similarly, we analyzed the three-dimensional Ising model and the XY models (Table 2), and the three-dimensional Heisenberg model (Table 3). In the latter cases, our results agree, within their precision, with previous literature, though we also discussed discrepancies, such as the topological transition in the XY 2D model.


In summary, this and prior work~\cite{Lima24,Lima25,Carrasco26} provide strong evidence that the mean-field fractal formulation for the correlation function (\ref{G3}) is effective for describing equilibrium symmetry-breaking phase transitions.

However, our framework does not capture topological phase transitions in the two-dimensional XY model, indicating a need for further investigation to address this limitation.
Another relevant question concerns the underlying geometric object with fractal dimension \(d_R\).


Since Eq.~(\ref{eta}) was derived for a general continuous order parameter $\phi(\vec{r})$, it is reasonable to anticipate its extension to non-equilibrium dynamics.
Exploring these directions~\cite{Henkel26,Ziff86,GomesFilho22}, especially in dynamic phase transitions~\cite{Ziff86,Pinto16,Pinto17,Santos24,Li25,Gutierrez26}, requires including temporal dependence in the autocorrelation function (\ref{Cdef}). This will open new avenues for understanding the relationship between critical phenomena, fractal geometry, and universality.


Recent research related to this work, such as Schr\"{o}dinger invariance in the voter model~\cite{Henkel26}, conformal scaling~\cite{Weberszpil26}, and Fisher curvature scaling at critical points~\cite{Zhuravlev26}, illustrates the field's promise and open questions.


\section*{Acknowledgements}

The authors acknowledge the Coordena\c{c}\~ao de Aperfei\c{c}oamento de Pessoal de N\'ivel Superior - CAPES.
FAO acknowledges financial support from Funda\c{c}\~ao de Amparo \`a Ci\^encia e Tecnologia de Pernambuco - FACEPE and CNPq (Grants APV-0064-1.05/25 and 303119/2022-5).
ALRB acknowledges financial support from CNPq (Grants 302502/2025-4 and 406836/2022-1 INCT of Spintronics and Advanced Magnetic Nanostructures - SpinNanoMag).

\bibliography{ReferencesV2}

@article{Alves16,
  title = {Analysis of etching at a solid-solid interface},
  author = {Alves, Washington S. and Rodrigues, Evandro A. and Fernandes, Henrique A. and Mello, Bernardo A. and Oliveira, Fernando A. and Costa, Ismael V. L.},
  journal = {Phys. Rev. E},
  volume = {94},
  issue = {4},
  pages = {042119},
  numpages = {7},
  year = {2016},
  month = {Oct},
  publisher = {American Physical Society},
  doi = {10.1103/PhysRevE.94.042119},
  url = {https://link.aps.org/doi/10.1103/PhysRevE.94.042119}
}

@ARTICLE{Anjos21,

AUTHOR={dos Anjos, Petrus H. R.  and Gomes-Filho, M\xc3\xa1rcio S.  and Alves, Washington S.  and Azevedo, David L.  and Oliveira, Fernando A. },

TITLE={The Fractal Geometry of Growth: Fluctuation\xe2\x80\x93Dissipation Theorem and Hidden Symmetry},

JOURNAL={Frontiers in Physics},

VOLUME={9},

pages = {741590},

YEAR={2021},

URL={https://www.frontiersin.org/journals/physics/articles/10.3389/fphy.2021.741590},

DOI={10.3389/fphy.2021.741590},

ISSN={2296-424X}}

@article{Averin10,
  title = {Violation of the Fluctuation-Dissipation Theorem in Time-Dependent Mesoscopic Heat Transport},
  author = {Averin, Dmitri V. and Pekola, Jukka P.},
  journal = {Phys. Rev. Lett.},
  volume = {104},
  issue = {22},
  pages = {220601},
  numpages = {4},
  year = {2010},
  month = {Jun},
  publisher = {American Physical Society},
  doi = {10.1103/PhysRevLett.104.220601},
  url = {https://link.aps.org/doi/10.1103/PhysRevLett.104.220601}
}

@book{Barabasi95,
title={Fractal concepts in surface growth},
author={Barabasi, A-L and Stanley, Harry Eugene and others},
year={1995},
publisher={Cambridge university press}
}

@article{Barrat98,
  title = {Monte Carlo simulations of the violation of the fluctuation-dissipation theorem in domain growth processes},
  author = {Barrat, A.},
  journal = {Phys. Rev. E},
  volume = {57},
  issue = {3},
  pages = {3629--3632},
  numpages = {0},
  year = {1998},
  month = {Mar},
  publisher = {American Physical Society},
  doi = {10.1103/PhysRevE.57.3629},
  url = {https://link.aps.org/doi/10.1103/PhysRevE.57.3629}
}

@article{Bellon02,
title = {Experimental study of the fluctuation dissipation relation during an aging process},
journal = {Physica D: Nonlinear Phenomena},
volume = {168},
pages = {325-335},
year = {2002},
issn = {0167-2789},
doi = {https://doi.org/10.1016/S0167-2789(02)00520-1},
url = {https://www.sciencedirect.com/science/article/pii/S0167278902005201},
author = {L Bellon and S Ciliberto}
}

@incollection{Bellon06,
  title={Thermal noise properties of two aging materials},
  author={Bellon, Ludovic and Buisson, Lionel and Ciccotti, Matteo and Ciliberto, Sergio and Douarche, Fr{\\\\e}d{\\\\e}ric},
  booktitle={Jamming, Yielding, and Irreversible Deformation in Condensed Matter},
  pages={23--52},
  year={2006},
  publisher={Springer}
}

@article{Brezin85,
  title={Finite size effects in phase transitions},
  author={Br{\'e}zin, E and Zinn-Justin, J},
  journal={Nuclear Physics B},
  volume={257},
  pages={867--893},
  year={1985},
  publisher={Elsevier}
}

@article{Campostrini02,
  title={Critical exponents and equation of state of the three-dimensional Heisenberg universality class},
  author={Campostrini, Massimo and Hasenbusch, Martin and Pelissetto, Andrea and Rossi, Paolo and Vicari, Ettore},
  journal={Physical Review B},
  volume={65},
  number={14},
  pages={144520},
  year={2002},
  publisher={APS}
}

@book{Cardy96,
  title={Scaling and renormalization in statistical physics},
  author={Cardy, John},
  volume={5},
  year={1996},
  publisher={Cambridge university press}
}

@Article{Carrasco26,
author={Carrasco, Ismael S. S.
and de Lima, Henrique A.
and Oliveira, Fernando A.},
title={A new fractal mean-field analysis in phase transition},
journal={The European Physical Journal Special Topics},
year={2026},
month={Apr},
day={28},
issn={1951-6401},
doi={10.1140/epjs/s11734-026-02327-6},
url={https://doi.org/10.1140/epjs/s11734-026-02327-6}
}

@article{Chester20,
  title={Carving out OPE space and precise O (2) model critical exponents},
  author={Chester, Shai M and Landry, Walter and Liu, Junyu and Poland, David and Simmons-Duffin, David and Su, Ning and Vichi, Alessandro},
  journal={Journal of High Energy Physics},
  volume={2020},
  number={6},
  pages={1--52},
  year={2020},
  publisher={Springer}
}

@article{Coniglio89,
  title = {Fractal structure of Ising and Potts clusters: Exact results},
  author = {Coniglio, Antonio},
  journal = {Phys. Rev. Lett.},
  volume = {62},
  issue = {26},
  pages = {3054--3057},
  numpages = {0},
  year = {1989},
  month = {Jun},
  publisher = {American Physical Society},
  doi = {10.1103/PhysRevLett.62.3054},
  url = {https://link.aps.org/doi/10.1103/PhysRevLett.62.3054}
}

@article{Costa03,
doi = {10.1209/epl/i2003-00514-3},
url = {https://dx.doi.org/10.1209/epl/i2003-00514-3},
year = {2003},
month = {jul},
publisher = {},
volume = {63},
number = {2},
pages = {173},
author = {I. V. L. Costa and  R. Morgado and  M. V. B. T. Lima and  F. A. Oliveira},
title = {The Fluctuation-Dissipation Theorem fails for fast superdiffusion},
journal = {Europhysics Letters}
}

@article{Costa06,
title = {Mixing, ergodicity and slow relaxation phenomena},
journal = {Physica A: Statistical Mechanics and its Applications},
volume = {371},
number = {1},
pages = {130-134},
year = {2006},
issn = {0378-4371},
doi = {https://doi.org/10.1016/j.physa.2006.04.096},
url = {https://www.sciencedirect.com/science/article/pii/S0378437106005413},
author = {I.V.L. Costa and M.H. Vainstein and L.C. Lapas and A.A. Batista and F.A. Oliveira}
}

@article{Crisanti03,
doi = {10.1088/0305-4470/36/21/201},
url = {https://dx.doi.org/10.1088/0305-4470/36/21/201},
year = {2003},
month = {may},
publisher = {},
volume = {36},
number = {21},
pages = {R181},
author = {A Crisanti and  F Ritort},
title = {Violation of the fluctuation\\xe2\\x80\\x93dissipation theorem in glassy systems: basic notions and the numerical evidence},
journal = {Journal of Physics A: Mathematical and General}
}

@article{Cruz23,
  title={Percolation on fractal networks: A survey},
  author={Cruz, Miguel Angel Martinez and Ortiz, Juli{\\a}n Pati{\\~n}o and Ortiz, Miguel Pati{\\~n}o and Balankin, Alexander},
  journal={Fractal and Fractional},
  volume={7},
  number={3},
  pages={231},
  year={2023},
  publisher={MDPI}
}

@article{Devakul19,
  title={{Fractal Symmetric Phases of Matter}},
  author={Trithep Devakul and Yizhi You and F. J. Burnell and S. L. Sondhi},
  journal={SciPost Phys.},
  volume={6},
  pages={007},
  year={2019},
  publisher={SciPost},
  doi={10.21468/SciPostPhys.6.1.007},
  url={https://scipost.org/10.21468/SciPostPhys.6.1.007},
}

@article{Family85,
  title={Scaling of the active zone in the Eden process on percolation networks and the ballistic deposition model},
  author={Family, Fereydoon and Vicsek, Tamas},
  journal={Journal of Physics A: Mathematical and General},
  volume={18},
  number={2},
  pages={L75},
  year={1985},
  publisher={IOP Publishing}
}

@article{f.y.wu,
  title = {The Potts model},
  author = {Wu, F. Y.},
  journal = {Rev. Mod. Phys.},
  volume = {54},
  issue = {1},
  pages = {235--268},
  numpages = {0},
  year = {1982},
  month = {Jan},
  publisher = {American Physical Society},
  doi = {10.1103/RevModPhys.54.235},
  url = {https://link.aps.org/doi/10.1103/RevModPhys.54.235}
}

@article{Fisher64,
title={Correlation functions and the critical region of simple fluids },
author={Fisher, Michael E.},
journal={Journal of Mathematical Physics},
volume={5},
number={7},
pages={944--962},
year={1964},
doi={10.1063/1.1704197}
}

@article{Gong19,
doi = {10.1088/0256-307X/36/7/076801},
url = {https://doi.org/10.1088/0256-307X/36/7/076801},
year = {2019},
month = {jun},
publisher = {Chinese Physical Society and IOP Publishing Ltd},
volume = {36},
number = {7},
pages = {076801},
author = {Gong, Yan and Guo, Jingwen and Li, Jiaheng and Zhu, Kejing and Liao, Menghan and Liu, Xiaozhi and Zhang, Qinghua and Gu, Lin and Tang, Lin and Feng, Xiao and Zhang, Ding and Li, Wei and Song, Canli and Wang, Lili and Yu, Pu and Chen, Xi and Wang, Yayu and Yao, Hong and Duan, Wenhui and Xu, Yong and Zhang, Shou-Cheng and Ma, Xucun and Xue, Qi-Kun and He, Ke},
title = {Experimental Realization of an Intrinsic Magnetic Topological Insulator*},
journal = {Chinese Physics Letters}
}

@book{Goldenfeld18,
  title={Lectures on phase transitions and the renormalization group},
  author={Goldenfeld, Nigel},
  year={2018},
  publisher={CRC Press}
}

@article{Gomes19,
title={From cellular automata to growth dynamics: The Kardar-Parisi-Zhang universality class},
author={Gomes, Waldenor P and Penna, Andr{\\\\e} LA and Oliveira, Fernando A},
journal={Physical Review E},
volume={100},
number={2},
pages={020101},
year={2019},
publisher={APS}
}

@article{GomesFilho21,
doi = {10.1209/0295-5075/133/10001},
url = {https://dx.doi.org/10.1209/0295-5075/133/10001},
year = {2021},
month = {mar},
publisher = {EDP Sciences, IOP Publishing and Societ\\xc3\\xa0 Italiana di Fisica},
volume = {133},
number = {1},
pages = {10001},
author = {M\\xc3\\xa1rcio S. Gomes-Filho and Fernando A. Oliveira},
title = {The hidden fluctuation-dissipation theorem for growth(a)},
journal = {Europhysics Letters}
}

@article{GomesFilho22,
  title = {Modeling the diffusion-erosion crossover dynamics in drug release},
  author = {Gomes-Filho, M\'arcio Sampaio and Oliveira, Fernando Albuquerque and Barbosa, Marco Aur\'elio Alves},
  journal = {Phys. Rev. E},
  volume = {105},
  issue = {4},
  pages = {044110},
  numpages = {12},
  year = {2022},
  month = {Apr},
  publisher = {American Physical Society},
  doi = {10.1103/PhysRevE.105.044110},
  url = {https://link.aps.org/doi/10.1103/PhysRevE.105.044110}
}

@Article{GomesFilho24,
AUTHOR = {Gomes-Filho, M\xc3\xa1rcio S. and de Castro, Pablo and Liarte, Danilo B. and Oliveira, Fernando A.},
TITLE = {Restoring the Fluctuation\xe2\x80\x93Dissipation Theorem in Kardar\xe2\x80\x93Parisi\xe2\x80\x93Zhang Universality Class through a New Emergent Fractal Dimension},
JOURNAL = {Entropy},
VOLUME = {26},
YEAR = {2024},
NUMBER = {3},
pages = {260},
ARTICLE-NUMBER = {260},
URL = {https://www.mdpi.com/1099-4300/26/3/260},
ISSN = {1099-4300},
DOI = {10.3390/e26030260}
}

@Article{GomesFilho25,
title = {The fluctuation–dissipation relations: Growth, diffusion, and beyond},
journal = {Physics Reports},
volume = {1141},
pages = {1-43},
year = {2025},
issn = {0370-1573},
doi = {https://doi.org/10.1016/j.physrep.2025.07.004},
url = {https://www.sciencedirect.com/science/article/pii/S037015732500198X},
author = {Márcio Sampaio Gomes-Filho and Luciano Calheiros Lapas and Ewa Gudowska-Nowak and Fernando Albuquerque Oliveira}
}

@article{Grigera99,
  title = {Observation of Fluctuation-Dissipation-Theorem Violations in a Structural Glass},
  author = {Grigera, Tom\\\\as S. and Israeloff, N. E.},
  journal = {Phys. Rev. Lett.},
  volume = {83},
  issue = {24},
  pages = {5038--5041},
  numpages = {0},
  year = {1999},
  month = {Dec},
  publisher = {American Physical Society},
  doi = {10.1103/PhysRevLett.83.5038},
  url = {https://link.aps.org/doi/10.1103/PhysRevLett.83.5038}
}

@book{Grimmett06,
  title={The random-cluster model},
  author={Grimmett, Geoffrey},
  volume={333},
  year={2006},
  publisher={Springer}
}

@article{Guillou80,
  title={Critical exponents from field theory},
  author={Le Guillou, JC and Zinn-Justin, Jean},
  journal={Physical Review B},
  volume={21},
  number={9},
  pages={3976},
  year={1980},
  publisher={APS}
}

@article{Gutierrez26,
  title={Dynamical phase diagram of synchronization in one dimension: universal behavior from Edwards-Wilkinson to random deposition through Kardar-Parisi-Zhang},
  author={Gutierrez, Ricardo and Cuerno, Rodolfo},
  journal={arXiv preprint arXiv:2604.06040},
  year={2026}
}

@article{Marcos26,
  title={On the upper critical dimension of the KPZ universality class: KPZ and related equations on a fully connected graph},
  author={Marcos, JM and Mel{\'e}ndez, JJ and Cuerno, R and Ruiz-Lorenzo, JJ},
  journal={arXiv preprint arXiv:2603.02000},
  year={2026}
}

@article{Hayashi07,
title={Violation of the fluctuation-dissipation theorem in a protein system},
author={Hayashi, Kumiko and Takano, Mitsunori},
journal={Biophysical journal},
volume={93},
number={3},
pages={895--901},
year={2007},
publisher={Elsevier},
doi ={10.1529/biophysj.106.100487},
        url= {https://doi.org/10.1529/biophysj.106.100487}

}

@article{Henkel26,
  title={Schr{\"o}dinger-invariance in the voter model},
  author={Henkel, Malte and Stoimenov, Stoimen},
  journal={International Journal of Theoretical Physics},
  volume={65},
  number={2},
  pages={48},
  year={2026},
  publisher={Springer}
}

@article{Holm93,
  title={Critical exponents of the classical three-dimensional Heisenberg model: A single-cluster Monte Carlo study},
  author={Holm, Christian and Janke, Wolfhard},
  journal={Physical Review B},
  volume={48},
  number={2},
  pages={936},
  year={1993},
  publisher={APS}
}

@article{Kardar86,
  title = {Dynamic Scaling of Growing Interfaces},
  author = {Kardar, Mehran and Parisi, Giorgio and Zhang, Yi-Cheng},
  journal = {Phys. Rev. Lett.},
  volume = {56},
  issue = {9},
  pages = {889--892},
  numpages = {0},
  year = {1986},
  month = {Mar},
  publisher = {American Physical Society},
  doi = {10.1103/PhysRevLett.56.889},
  url = {https://link.aps.org/doi/10.1103/PhysRevLett.56.889}
}

@article{Kosterlitz73,
  title={Ordering, metastability and phase transitions in two-dimensional systems},
  author={Kosterlitz, John Michael and Thouless, David James},
  journal={Journal of Physics C: Solid State Physics},
  volume={6},
  number={7},
  pages={1181--1203},
  year={1973}
}

@article{Kosterlitz74,
  title={The critical properties of the two-dimensional xy model},
  author={Kosterlitz, J Michael},
  journal={Journal of Physics C: Solid State Physics},
  volume={7},
  number={6},
  pages={1046--1060},
  year={1974}
}

@article{Kosterlitz17nobel,
  title={Nobel lecture: Topological defects and phase transitions},
  author={Kosterlitz, John Michael},
  journal={Reviews of Modern Physics},
  volume={89},
  number={4},
  pages={040501},
  year={2017},
  publisher={APS}
}

@article{Kroger00,
  title={Fractal geometry in quantum mechanics, field theory and spin systems},
  author={Kroger, Helmut},
  journal={Physics Reports},
  volume={323},
  number={2},
  pages={81--181},
  year={2000},
  publisher={Elsevier}
}

@article{Lapas07,
doi = {10.1209/0295-5075/77/37004},
url = {https://dx.doi.org/10.1209/0295-5075/77/37004},
year = {2007},
month = {jan},
publisher = {},
volume = {77},
number = {3},
pages = {37004},
author = {L. C. Lapas and I. V. L. Costa and M. H. Vainstein and F. A. Oliveira},
title = {Entropy, non-ergodicity and non-Gaussian behaviour in ballistic transport},
journal = {Europhysics Letters}
}

@article{Lapas08,
  title = {Khinchin Theorem and Anomalous Diffusion},
  author = {Lapas, Luciano C. and Morgado, Rafael and Vainstein, Mendeli H. and Rub\\i, J. Miguel and Oliveira, Fernando A.},
  journal = {Phys. Rev. Lett.},
  volume = {101},
  issue = {23},
  pages = {230602},
  numpages = {4},
  year = {2008},
  month = {Dec},
  publisher = {American Physical Society},
  doi = {10.1103/PhysRevLett.101.230602},
  url = {https://link.aps.org/doi/10.1103/PhysRevLett.101.230602}
}

@article{Li25,
  title={Modeling Phase Transitions in Starling Flocks Using Fractal Dimension of Self-Affine Functions},
  author={Li, Kunyuan and Zhang, Xiongwei and Yao, Kui and Zhang, Kai and Sun, Meng and He, Ming and Liu, Kefeng and Wang, Yangjun},
  journal={Fractal and Fractional},
  volume={10},
  number={1},
  pages={17},
  year={2025},
  publisher={MDPI}
}

@article{Lima24,
   title={Geometrical interpretation of critical exponents},
   volume={110},
   ISSN={2470-0053},
   url={http://dx.doi.org/10.1103/PhysRevE.110.L062107},
   DOI={10.1103/physreve.110.l062107},
   number={6},
   journal={Physical Review E},
   publisher={American Physical Society (APS)},
   author={Lima, Henrique A. and Luis, Edwin E. Mozo and Carrasco, Ismael S. S. and Hansen, Alex and Oliveira, Fernando A.},
   year={2024},
   month=dec }

@article{Lima25,
  title = {Scaling, fractal dynamics, and critical exponents: Application in a noninteger-dimensional Ising model},
  author = {de Lima, Henrique A. and Carrasco, Ismael S. S. and Santos, Marcio and Oliveira, Fernando A.},
  journal = {Phys. Rev. E},
  volume = {112},
  issue = {4},
  pages = {044109},
  numpages = {7},
  year = {2025},
  month = {Oct},
  publisher = {American Physical Society},
  doi = {10.1103/rh4r-7mfv},
  url = {https://link.aps.org/doi/10.1103/rh4r-7mfv}
}

@article{lima26,
  title={Strong universality class in disordered systems},
  author={Lima, Henrique A and Hermann, Kaue and Carrasco, Ismael SS and de Almeida, Jairo RL and Oliveira, Fernando A},
  journal={arXiv preprint arXiv:2605.15441},
  year={2026}
}

@article{Luis22,
doi = {10.1088/1742-5468/ac7e3f},
url = {https://dx.doi.org/10.1088/1742-5468/ac7e3f},
year = {2022},
month = {aug},
publisher = {IOP Publishing and SISSA},
volume = {2022},
number = {8},
pages = {083202},
author = {Edwin E Mozo Luis and Thiago A de Assis and Fernando A Oliveira},
title = {Unveiling the connection between the global roughness exponent and interface fractal dimension in EW and KPZ lattice models},
journal = {Journal of Statistical Mechanics: Theory and Experiment}
}

@article{Muslih10,
  title={Riesz fractional derivatives and fractional dimensional space},
  author={Muslih, Sami I and Agrawal, Om P},
  journal={International Journal of Theoretical Physics},
  volume={49},
  pages={270--275},
  year={2010},
  publisher={Springer},
  doi = {10.1007/s10773-009-0200-1},
  url = {https://doi.org/10.1007/s10773-009-0200-1}
}

@article{Nijs83,
  title={Extended scaling relations for the magnetic critical exponents of the Potts model},
  author={den Nijs, Marcel},
  journal={Physical Review B},
  volume={27},
  number={3},
  pages={1674},
  year={1983},
  publisher={APS}
}

@ARTICLE{Nowak22,
  
AUTHOR={Gudowska-Nowak, Ewa  and Oliveira, Fernando A.  and Wio, Horacio Sergio },
         
TITLE={Editorial: The Fluctuation-Dissipation Theorem Today},
        
JOURNAL={Frontiers in Physics},
        
VOLUME={Volume 10 - 2022},

YEAR={2022},

URL={https://www.frontiersin.org/journals/physics/articles/10.3389/fphy.2022.859799},

DOI={10.3389/fphy.2022.859799},

ISSN={2296-424X},

}

@article{Perez-Madrid09,
  title = {Heat Exchange between Two Interacting Nanoparticles beyond the Fluctuation-Dissipation Regime},
  author = {Perez-Madrid, Agustin and Lapas, Luciano Calheiros and Rub{\\i}, J. Miguel},
  journal = {Phys. Rev. Lett.},
  volume = {103},
  issue = {4},
  pages = {048301},
  numpages = {4},
  year = {2009},
  month = {Jul},
  publisher = {American Physical Society},
  doi = {10.1103/PhysRevLett.103.048301},
  url = {https://link.aps.org/doi/10.1103/PhysRevLett.103.048301}
}

@article{Pinto16,
  title = {Thermodynamics aspects of noise-induced phase synchronization},
  author = {Pinto, Pedro D. and Oliveira, Fernando A. and Penna, Andr\\e L. A.},
  journal = {Phys. Rev. E},
  volume = {93},
  issue = {5},
  pages = {052220},
  numpages = {10},
  year = {2016},
  month = {May},
  publisher = {American Physical Society},
  doi = {10.1103/PhysRevE.93.052220},
  url = {https://link.aps.org/doi/10.1103/PhysRevE.93.052220}
}

@article{Pinto17,
doi = {10.1209/0295-5075/117/50009},
url = {https://dx.doi.org/10.1209/0295-5075/117/50009},
year = {2017},
month = {may},
publisher = {EDP Sciences, IOP Publishing and Societ\xc3\xa0 Italiana di Fisica},
volume = {117},
number = {5},
pages = {50009},
author = {Pedro D. Pinto and Andr\xc3\xa9 L. A. Penna and Fernando A. Oliveira},
title = {Critical behavior of noise-induced phase synchronization},
journal = {Europhysics Letters}
}

@inproceedings{Potts52,
  title={Some generalized order-disorder transformations},
  author={Potts, Renfrey Burnard},
  booktitle={Mathematical proceedings of the cambridge philosophical society},
  volume={48},
  number={1},
  pages={106--109},
  year={1952},
  organization={Cambridge University Press}
}

@article{Ricci-Tersenghi00,
  title = {Two Time Scales and Violation of the Fluctuation-Dissipation Theorem in a Finite Dimensional Model for Structural Glasses},
  author = {Ricci-Tersenghi, Federico and Stariolo, Daniel A. and Arenzon, Jeferson J.},
  journal = {Phys. Rev. Lett.},
  volume = {84},
  issue = {19},
  pages = {4473--4476},
  numpages = {0},
  year = {2000},
  month = {May},
  publisher = {American Physical Society},
  doi = {10.1103/PhysRevLett.84.4473},
  url = {https://link.aps.org/doi/10.1103/PhysRevLett.84.4473}
}

@article{Rodrigues15,
doi = {10.1088/1751-8113/48/3/035001},
url = {https://dx.doi.org/10.1088/1751-8113/48/3/035001},
year = {2015},
month = {dec},
publisher = {IOP Publishing},
volume = {48},
number = {3},
pages = {035001},
author = {Evandro A Rodrigues and Bernardo A Mello and Fernando A Oliveira},
title = {Growth exponents of the etching model in high dimensions},
journal = {Journal of Physics A: Mathematical and Theoretical}
}

@article{Rodrigues24,
doi = {10.1088/1742-5468/ad1d57},
url = {https://dx.doi.org/10.1088/1742-5468/ad1d57},
year = {2024},
month = {jan},
publisher = {IOP Publishing},
volume = {2024},
number = {1},
pages = {013209},
author = {Evandro A Rodrigues and Edwin E Mozo Luis and Thiago A de Assis and Fernando A Oliveira},
title = {Universal scaling relations for growth phenomena},
journal = {Journal of Statistical Mechanics: Theory and Experiment}
}

@article{Rodriguez19,
  title = {Stochastic entropies and fluctuation theorems for a discrete one-dimensional Kardar-Parisi-Zhang system},
  author = {Rodriguez, Miguel A. and Wio, Horacio S.},
  journal = {Phys. Rev. E},
  volume = {100},
  issue = {3},
  pages = {032111},
  numpages = {6},
  year = {2019},
  month = {Sep},
  publisher = {American Physical Society},
  doi = {10.1103/PhysRevE.100.032111},
  url = {https://link.aps.org/doi/10.1103/PhysRevE.100.032111}
}

@article{Santos24,
  title = {Phase transitions in the Ziff-Gulari-Barshad model operating on periodic conditions},
  author = {Santos, M. and Oliveira, Fernando A.},
  journal = {Phys. Rev. E},
  volume = {110},
  issue = {4},
  pages = {044122},
  numpages = {9},
  year = {2024},
  month = {Oct},
  publisher = {American Physical Society},
  doi = {10.1103/PhysRevE.110.044122},
  url = {https://link.aps.org/doi/10.1103/PhysRevE.110.044122}
}

@article{Suzuki8,
  title={Phase transition and fractals},
  author={Suzuki, Masuo},
  journal={Progress of Theoretical Physics},
  volume={69},
  number={1},
  pages={65--76},
  year={1983},
  publisher={Oxford University Press}
}

@article{Vainstein05,
title = {Stochastic description of the dynamics of a random-exchange Heisenberg chain},
journal = {Physics Letters A},
volume = {339},
number = {1},
pages = {33-38},
year = {2005},
issn = {0375-9601},
doi = {https://doi.org/10.1016/j.physleta.2005.02.059},
url = {https://www.sciencedirect.com/science/article/pii/S0375960105003014},
author = {M.H. Vainstein and R. Morgado and F.A. Oliveira and F.A.B.F. {de Moura} and M.D. Coutinho-Filho}
}

@article{Wen23,
  title={Ergodic Measure and Potential Control of Anomalous Diffusion},
  author={Wen, Bao and Li, Ming-Gen and Liu, Jian and Bao, Jing-Dong},
  journal={Entropy},
  volume={25},
  number={7},
  pages={1012},
  year={2023},
  publisher={MDPI}
}

@article{Xu25,
  title={Correction-to-scaling exponent for percolation and the Fortuin-Kasteleyn Potts model in two dimensions},
  author={Xu, Yihao and Chen, Tao and Zhou, Zongzheng and Salas, Jes{\\u}s and Deng, Youjin},
  journal={Physical Review E},
  volume={111},
  number={3},
  pages={034108},
  year={2025},
  publisher={APS}
}

@article{Weberszpil26,
  title={Conformable scaling and critical phenomena: a unified framework for phase transitions},
  author={Weberszpil, Jos{\'e} and Metzler, Ralf},
  journal={Journal of Physics A: Mathematical and Theoretical},
  volume={59},
  number={8},
  pages={085001},
  year={2026},
  publisher={IOP Publishing}
}

@article{Zhuravlev26,
  title={Fisher Curvature Scaling at Critical Points: An Exact Information-Geometric Exponent from Periodic Boundary Conditions},
  author={Zhuravlev, Max},
  journal={arXiv preprint arXiv:2603.07651},
  year={2026}
}

@article{Ziff86,
  title={Kinetic phase transitions in an irreversible surface-reaction model},
  author={Ziff, Robert M and Gulari, Erdagon and Barshad, Yoav},
  journal={Physical review letters},
  volume={56},
  number={24},
  pages={2553},
  year={1986},
  publisher={APS}
}

@misc{dupuis2026rg,
  title={Renormalization group and critical phenomena},
  author={Dupuis, Nicolas},
  year={2026},
  howpublished={Lecture notes at LPTMC, Sorbonne Universit{\'e}},
  url={https://www.lptmc.jussieu.fr/user/dupuis/toc_short.pdf}
}

@article{Coniglio_1980,
doi = {10.1088/0305-4470/13/8/025},
url = {https://doi.org/10.1088/0305-4470/13/8/025},
year = {1980},
month = {aug},
publisher = {},
volume = {13},
number = {8},
pages = {2775},
author = {A Coniglio and W Klein},
title = {Clusters and Ising critical droplets: a renormalisation group approach},
journal = {Journal of Physics A: Mathematical and General}
}

@article{FORTUIN1972536,
title = {On the random-cluster model: I. Introduction and relation to other models},
journal = {Physica},
volume = {57},
number = {4},
pages = {536-564},
year = {1972},
issn = {0031-8914},
doi = {https://doi.org/10.1016/0031-8914(72)90045-6},
url = {https://www.sciencedirect.com/science/article/pii/0031891472900456},
author = {C.M. Fortuin and P.W. Kasteleyn}
}

@book{landau2021guide,
  title={A Guide to Monte Carlo Simulations in Statistical Physics},
  author={Landau, D. and Binder, K.},
  isbn={9781108490146},
  lccn={2020021931},
  url={https://books.google.com.br/books?id=xW-EzQEACAAJ},
  year={2021},
  publisher={Cambridge University Press}
}

\end{document}